\documentclass[conference]{IEEEtran}
\IEEEoverridecommandlockouts
\usepackage{amsmath,amsfonts}
\usepackage{bbding}
\usepackage{algorithmic}
\usepackage{algorithm}
\usepackage{array}
\usepackage[caption=false,font=normalsize,labelfont=sf,textfont=sf]{subfig}
\usepackage{textcomp}
\usepackage{stfloats}
\usepackage{url}
\usepackage{verbatim}
\usepackage{graphicx}
\usepackage{CJKutf8}
\usepackage{balance}
\usepackage{wasysym}
\usepackage{siunitx}
\usepackage{booktabs}
\usepackage{color, xcolor}
\usepackage{tikz,xcolor}
\usepackage{multirow}
\usepackage{caption}
\usepackage{cuted}
\usepackage[numbers,sort&compress]{natbib} 
\usepackage{hyperref}

\definecolor{lime}{HTML}{A6CE39}
\DeclareRobustCommand{\orcidicon}{
\begin{tikzpicture}
\draw[lime, fill=lime] (0,0)
circle[radius=0.16]
node[white]{{\fontfamily{qag}\selectfont \tiny \.{I}D}};
\end{tikzpicture}
\hspace{-2mm}
}
\foreach \x in {A, ..., Z}{%
\expandafter\xdef\csname orcid\x\endcsname{\noexpand\href{https://orcid.org/\csname orcidauthor\x\endcsname}{\noexpand\orcidicon}}
}

\graphicspath{{./figure/}}    

\begin{document}

\bstctlcite{IEEEexample:BSTcontrol}

\title{NeuroPDE: A Neuromorphic PDE Solver Based on Spintronic and Ferroelectric  Devices
\thanks{This work was supported by the National Key R\&D Program of China(2022YFB2803405) and the NSFC(62304257). ($^{\ast}$Corresponding author: Lizhou Wu and Tiejun Li.)}}

\author{\IEEEauthorblockN{Siqing Fu\hspace{-1.5mm}\orcidA{}, Lizhou Wu$^{\ast}$\hspace{-1.5mm}\orcidC, Tiejun Li$^{\ast}$\hspace{-1.5mm}\orcidH{}, Chunyuan Zhang\hspace{-1.5mm}\orcidD, Sheng Ma\hspace{-1.5mm}\orcidE, \\Jianmin Zhang\hspace{-1.5mm}\orcidG{}, Yuhan Tang\hspace{-1.5mm}\orcidF and Jixuan Tang\hspace{-1.5mm}\orcidI{}}
\IEEEauthorblockA{
\textit{College of Computer Science and Technology, National University of Defense Technology}\\
Changsha, China \\
\{fusiqingnudt, lizhou.wu, tjli, cyzhang, masheng, jmzhang, tangyuhan, tangjixuan19\}@nudt.edu.cn}
}

\maketitle

\begin{abstract}
In recent years, new methods for solving partial differential equations (PDEs) such as  Monte Carlo random walk methods have gained considerable attention. However, due to the lack of hardware-intrinsic randomness in the conventional von Neumann architecture, the performance of PDE solvers is limited. In this paper, we introduce NeuroPDE, a hardware design for neuromorphic PDE solvers that utilizes emerging spintronic and ferroelectric devices. NeuroPDE incorporates spin neurons that are capable of probabilistic transmission to emulate random walks, along with ferroelectric synapses that store continuous weights non-volatilely. The proposed NeuroPDE achieves a variance of less than 1e-2 compared to analytical solutions when solving diffusion equations, demonstrating a performance advantage of 3.48$\times$ to 315$\times$ speedup in execution time and an energy consumption advantage of 2.7$\times$ to 29.8$\times$ over advanced CMOS-based neuromorphic chips. By leveraging the inherent physical stochasticity of emerging devices, this study paves the way for future probabilistic neuromorphic computing systems.

\end{abstract}


\section{Introduction}
\label{Intro}

As a fundamental class of mathematical equations, \textit{partial differential equations} (PDEs)  express the relationship between unknown functions containing several variables and their partial derivatives with respect to  these variables. 
PDEs are widely used not only in all areas of physics but also in engineering 
to solve complex  problems such as wave propagation, heat conduction, fluid mechanics, and particle transport \cite{rasht2022physics, hu2020heat, gao2021bi, harada2020boltzmann}. 
In recent years, we have also seen a dramatic increase in the use of PDEs in computer sciences particularly for image processing and graphics. 
Due to the complexity of the equations, solving PDEs in an energy-efficient manner remains a big challenge and thus attracts great research attention. 

PDE solvers have evolved from classical analytical techniques \cite{henner2019partial} and numerical algorithms such as finite difference methods \cite{langtangen2017finite}, finite element methods \cite{Claes1990}, and spectral methods \cite{townsend2015automatic}, to recently neural networks methods \cite{blechschmidt2021three,karniadakis2021physics,meng2023pinn, Pestourie2023} and emerging \textit{Monte Carlo} (MC) random walk methods \cite{severa2018spiking, smith2020solving, smith2022neuromorphic,zhang2023monte}. Analytical and numerical methods struggle with solving high-dimensional and non-linear equations \cite{hu2024tackling}, whereas neural network methods tackle this challenge through a data-driven approach. However, neural network PDE solvers, which incorporate PDE constraints into training or utilize pre-generated data, can be computationally expensive to train \cite{Jiang2023}. Addressing these limitations, novel MC random walk PDE solvers \cite{zhang2023monte} employ a probabilistic PDE formulation to train unsupervised models using random particle walks, which can effectively solve convection-diffusion, Allen-Cahn, and Navier-Stokes equations. Additionally, Smith et al. \cite{smith2022neuromorphic} advanced random-walk-based PDE solving methodologies by implementing them on neuromorphic computing chips such as Loihi \cite{davies2018loihi} and TrueNorth \cite{Akopyan}, showing a substantial  improvement in energy consumption. Nevertheless, these CMOS-based neuromorphic chips lack inherent randomness \cite{misra2023probabilistic} and thus do not show a clear performance advantage over conventional CPUs and GPUs.

To enhance computational performance while reducing energy consumption, we present NeuroPDE, a neuromorphic PDE solver that utilizes spintronic \textit{magnetic tunnel junction} (MTJ) and \textit{ferroelectric tunnel junction} (FTJ) devices. This design seamlessly bridges the gap between the probabilistic MC random walk method and the underlying circuit design, exploiting the intrinsic physical randomness of MTJ devices to create a scalable neuromorphic architecture capable of performing random walk functions. Through both SPICE simulations at the circuit level and functional simulations at the system level, we evaluated the ability of NeuroPDE to solve steady-state diffusion equations, showing the solution variance between NeuroPDE and analytical solutions remains below 1e-2. Compared to CMOS-based neuromorphic PDE solvers \cite{smith2022neuromorphic} implemented on Loihi and TrueNorth\cite{davies2018loihi,Akopyan}, NeuroPDE demonstrates performance improvements of 3.48-315$\times$ and achieves power efficiency improvements  of 2.7-29.8$\times$.


The main contributions of this paper are as follows:
 
\begin{enumerate} 
    \item We design a synapse based on an FTJ device. Utilizing the memristive property of this device, the synapse non-volatilely stores continuous synaptic weights, enabling stable storage of Markov chain transition probabilities.
    
    \item We propose a neuron design based on an MTJ device. Using its probabilistic switching property, we achieve probabilistic activation, winner-takes-all, and self-inhibition, enabling simulation of random walks on a Markov chain through neural activation.
    
    \item Building on our neuron and synapse designs, we develop the neuromorphic PDE solver NeuroPDE. We perform a comprehensive circuit and system simulations of NeuroPDE to assess its accuracy, execution time, and energy consumption in solving PDEs. As an architectural exploration, NeuroPDE demonstrates the potential of hardware-integrated stochastic PDE solvers, paving the way for practical PDE hardware solutions.

\end{enumerate}

The structure of this paper is as follows: Section~\ref{Back} reviews the foundational concepts of MC random walks, and describes the novel devices used in the circuit construction. Section~\ref{Design} presents the circuit design of NeuroPDE, with a focus on the neuron and synapse circuits. Section~\ref{Experiments} details the experimental evaluations, including circuit-level and system-level simulations, and presents the performance and energy results. Section~\ref{Discussion} addresses the limitations and suggests directions for future research. Finally, Section~\ref{Conclusion} concludes the paper.

\section{Background}
\label{Back}

\subsection{MC Random Walk Method}
\label{Back:MC}

The MC random walk method \cite{MASUDA20171} relates particle trajectory simulations to PDE solutions by capturing stochastic behaviors, such as Brownian motion associated with thermal diffusion. This method involve spatial discretization, the construction of Markov chains to simulate particle movements, and MC sampling to approximate PDE solution spaces.

\begin{figure}[t]
\centering
\includegraphics[width=0.9\linewidth]{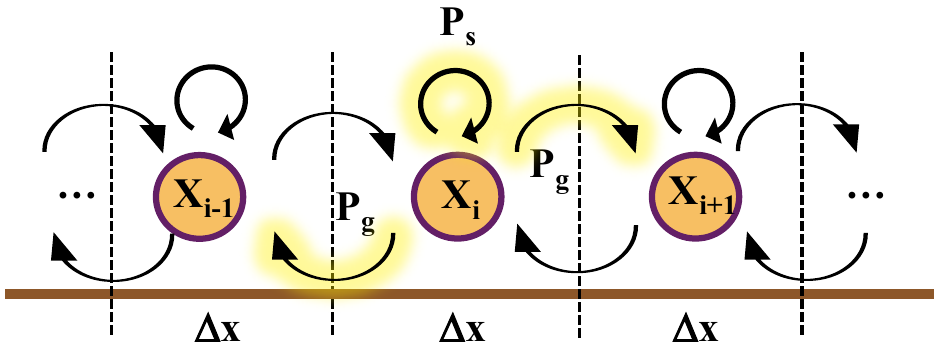}
\caption{Constructed Markov chain model for random walks.} 
\label{fig1}
\end{figure}

The Feynman-Kac formula \cite{Hu2019} extends the correspondence principle, suggesting a link between solutions of certain PDEs and their associated \textit{stochastic differential equations} (SDEs). Specifically, the PDEs of interest are those formulated as Equation (\ref{equPDE}):
\begin{equation}
\frac{\partial u}{\partial t}+\mu  (x,t)\frac{\partial u}{\partial x}+\frac{1}{2}\sigma ^{2}(x,t) \frac{\partial^2u}{\partial x^2}+f(x,t)u=0, 
\label{equPDE}
\end{equation}
where $u=u(x,t)$ is the function to be solved, $\mu(x,t)$ and $\sigma(x,t)$ represent the drift and diffusion terms, and $f(x,t)$ is a given function. The Feynman-Kac formula expresses the solution to the PDE via a stochastic process, as shown in Equation (\ref{equresolve}):
\begin{equation}
u(x,t)=\mathbb{E} [\phi(X_{t} )\mid  X_{t} =x] ,
\label{equresolve}
\end{equation}
where $X_{t}$ denotes Brownian motion at time $t$, which can be efficiently simulated using a Markov chain model. As depicted in Fig.~\ref{fig1}, we define a finite one-dimensional space with $N$ discrete positions as the state space $S=\left \{{\mathrm{X}_{1},\mathrm{X}_{2},...,\mathrm{X}_{i},...}  \right \} $, where each state represents the position of the walker in the one-dimensional space. The Markov chain is characterized by its state transition matrix $P$, specifying the probability of transitioning from state $i$ to state $j$: $P_{i,i+1}  =  P_{i,i-1}  =  P_{g}$, $P_{i,i}  =  P_{s}$. Boundary conditions are not detailed here, as they vary with the problem.

Smith et al. \cite{smith2022neuromorphic} provided a comprehensive elucidation on solving PDEs through MC random walks, alongside a discussion on its applicability to a spectrum of problems, including stock option pricing, Boltzmann flux density, the heat equation with dissipation, and simple beam bending, among others. Given that the adaptability of the method to various problems has already been deliberated upon, this paper exemplifies the hardware implementation efficacy of MC random walks and the feasibility of its extension across problem dimensions through two case studies: a steady-state thermal problem and a time-dependent thermal diffusion problem.

\subsection{MTJ and FTJ Device Basics}

\begin{figure}[t]
    \begin{minipage}[t]{0.65\linewidth}
        \centering
        \includegraphics[width=\textwidth]{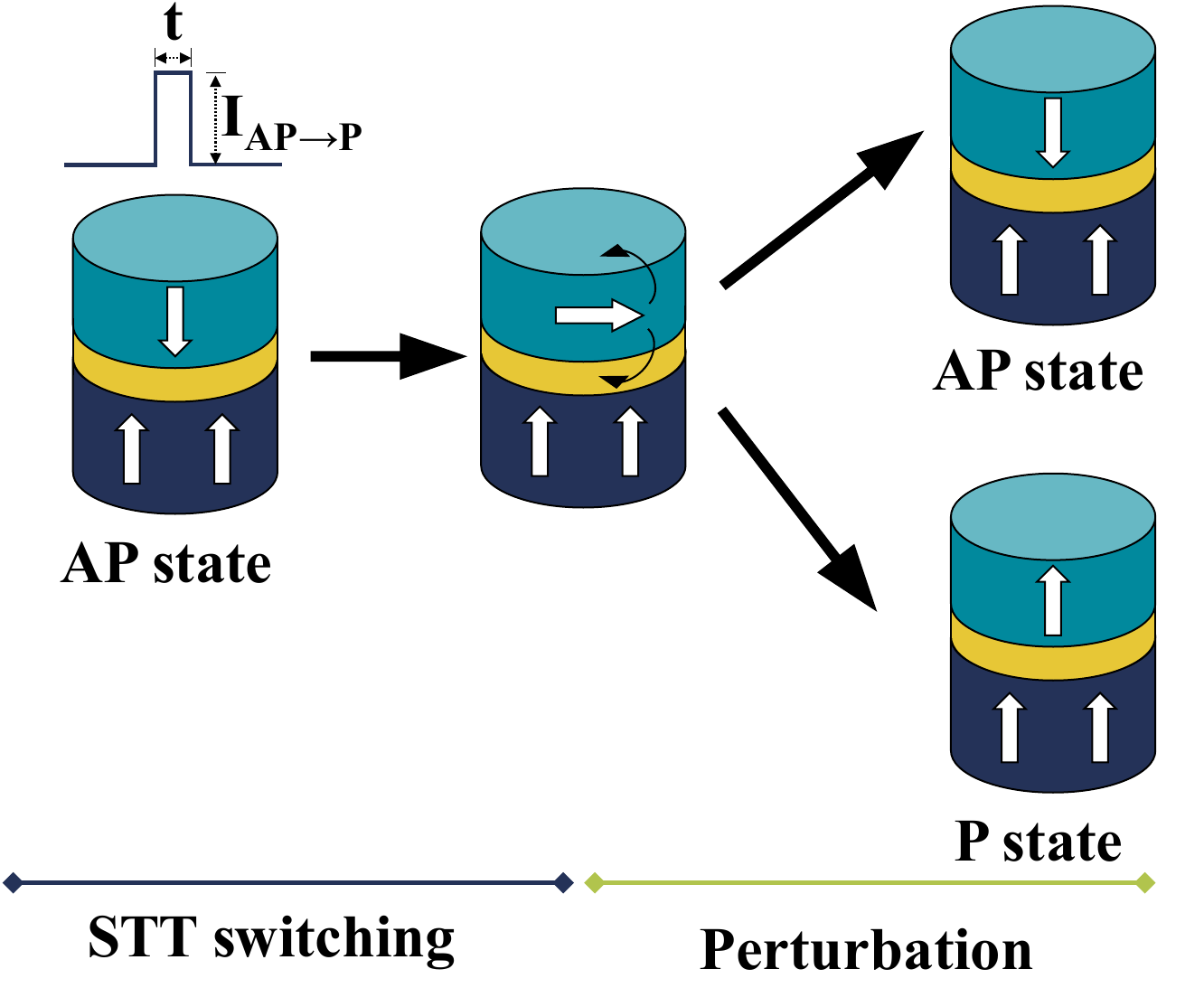} 
        \hfill 
        \centerline{\quad\quad\quad\quad\quad(a)}
    \end{minipage}%
    \begin{minipage}[t]{0.35\linewidth}
        \centering
        \includegraphics[width=\textwidth]{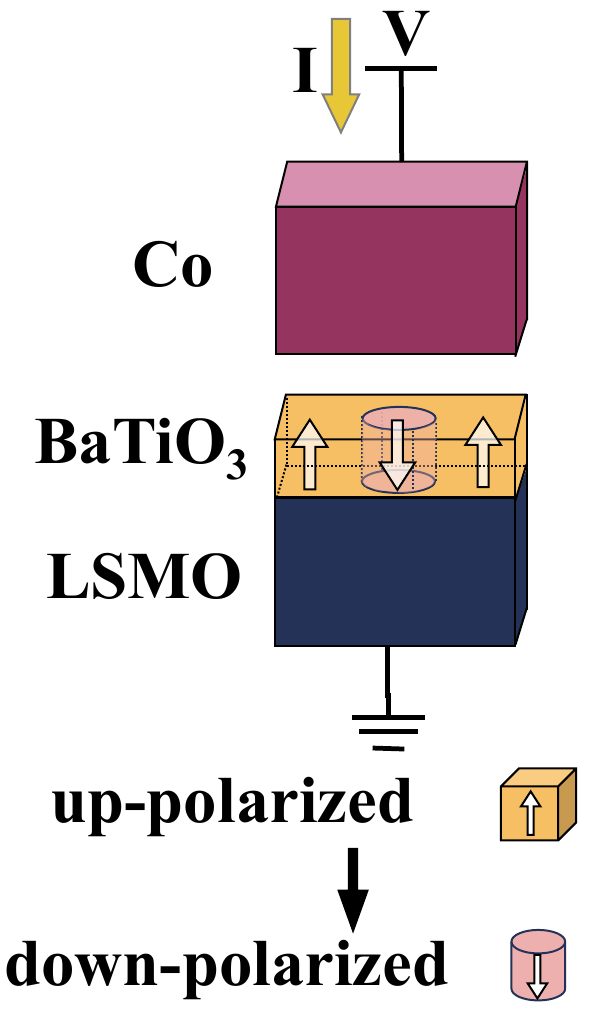}
        \centerline{(b)}
    \end{minipage}
    \caption{(a) STT switching and random thermal perturbations in STT-MTJs, (b) Programming voltage triggers down-polarized domain nucleation and propagation.}
    \label{fig2}
\end{figure}
\subsubsection{MTJ Device}
\label{Back:MTJ}

MTJs are widely studied emerging spintronic devices, fundamentally composed of a three-layer structure: CoFeB/MgO/CoFeB \cite{yamamoto2021perpendicular}. It includes a free layer with variable magnetization, a tunneling barrier, and a fixed layer with the constant magnetization.
Fig.~\ref{fig2}(a) depicts the stochastic magnetization switching process of an MTJ driven by limited \textit{spin-transfer torque} (STT) under thermal perturbation \cite{zhang2012}. In an initial \textit{antiparallel} (AP) state, a spin-polarized current $I_{\mathrm{AP}\rightarrow\mathrm{P}}$ perpendicular to the MTJ draws electrons from the fixed layer to the free layer. It exerts a torque known as STT on the magnetization of the free layer. The switching probability $P$  under a switching current with amplitude $I$ and duration $t$ is given by:
\begin{eqnarray} 
P(I,t) & = & 1-\exp\left(-\frac{t}{\tau}\right), \\
\tau(I) & = & \tau_{0}\exp\left[\Delta\left(1-\frac{I}{I_{c0}}\right)^{2}\right],
\label{equMTJ}
\end{eqnarray}
where $\tau$ represents the mean switching time, $\tau_{0}$ is the attempt time, $\Delta$ is the temperature-dependent thermal stability factor, and  $I_{c0}$ is the critical switching current at \SI{0}{\kelvin} \cite{Fukushima2014}. The MTJ in initial \textit{parallel} (P) state is probabilistically switched by a reverse current. By exploiting the physical randomness amplified by thermal perturbations in MTJs, we provide a stochastic source for neuromorphic systems simulating random walks \cite{Fu2023}.

\subsubsection{FTJ Device}
\label{Back:FTJ}

To store continuous weights, a device with variable resistance is essential. Recent research has identified memristive behavior in FTJs, where resistance varies continuously through ferroelectric polarization switching \cite{Garcia2014,max2020,wen2020,Fang2024}.

A typical FTJ structure comprises an ultrathin ferroelectric barrier and two electrode layers, as illustrated in Fig.~\ref{fig2}(b), with the composition Co/BaTiO$_3$(BTO)/LSMO \cite{Sandu2022}. Applying a voltage induces BTO ferroelectric polarization switching, modulating the barrier and altering electron tunneling probabilities, thus inducing resistive changes. From a perspective of ferroelectric domain dynamics, ferroelectric polarization switching in BTO, driven by domain wall nucleation and propagation, manifests as the FTJ resistance through the parallel resistances of opposing polarized domains.  In fully up-polarized ferroelectric films, a programming voltage initiates down-polarized domain nucleation and propagation, resulting in a continuous resistance alteration, as depicted in Fig.~\ref{fig2}(b).

\section{Proposed NeuroPDE Design}
\label{Design}


\subsection{ Design Philosophy and Architecture}

\begin{figure}[t]
\centering
\includegraphics[width=9cm]{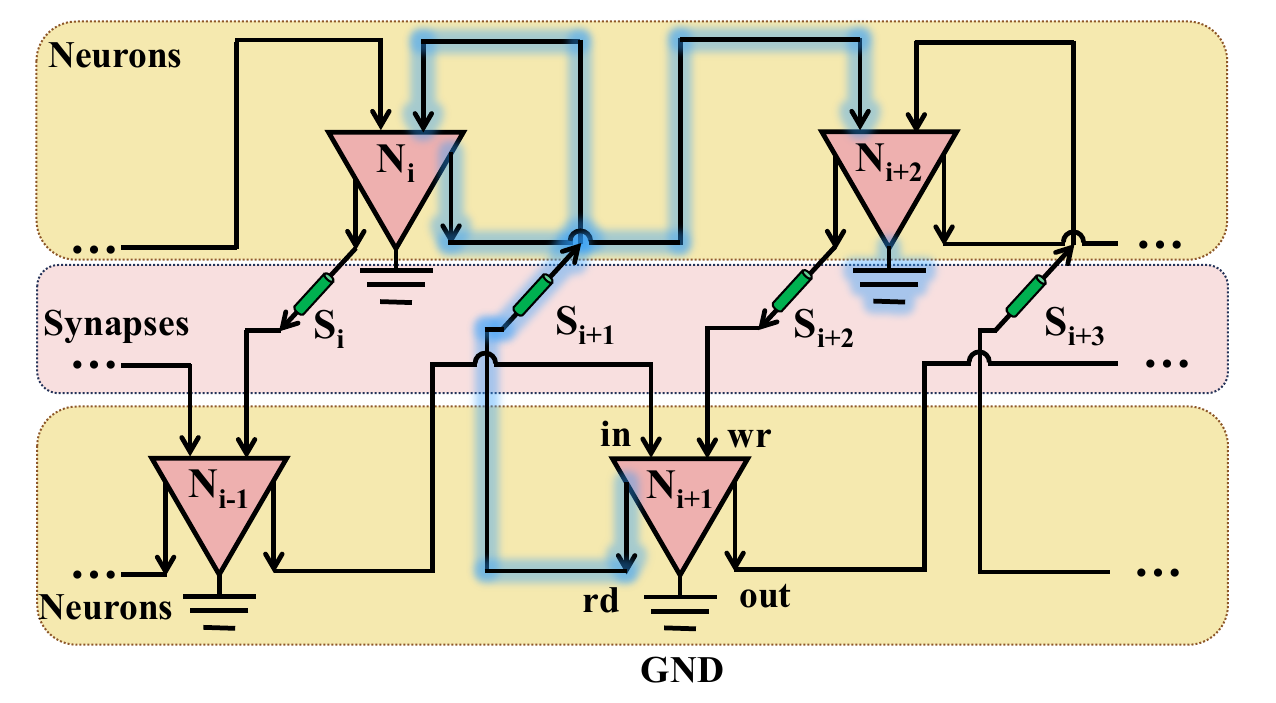}
\caption{NeuroPDE design overview: neurons arranged by parity and interconnected via synapses, highlighting activation pathways from neuron N$_{i+1}$ to neighboring neurons.} 
\label{fig4}
\vspace{-5pt} 
\end{figure}

\begin{figure}[t]
\centering
\includegraphics[width=7.5cm]{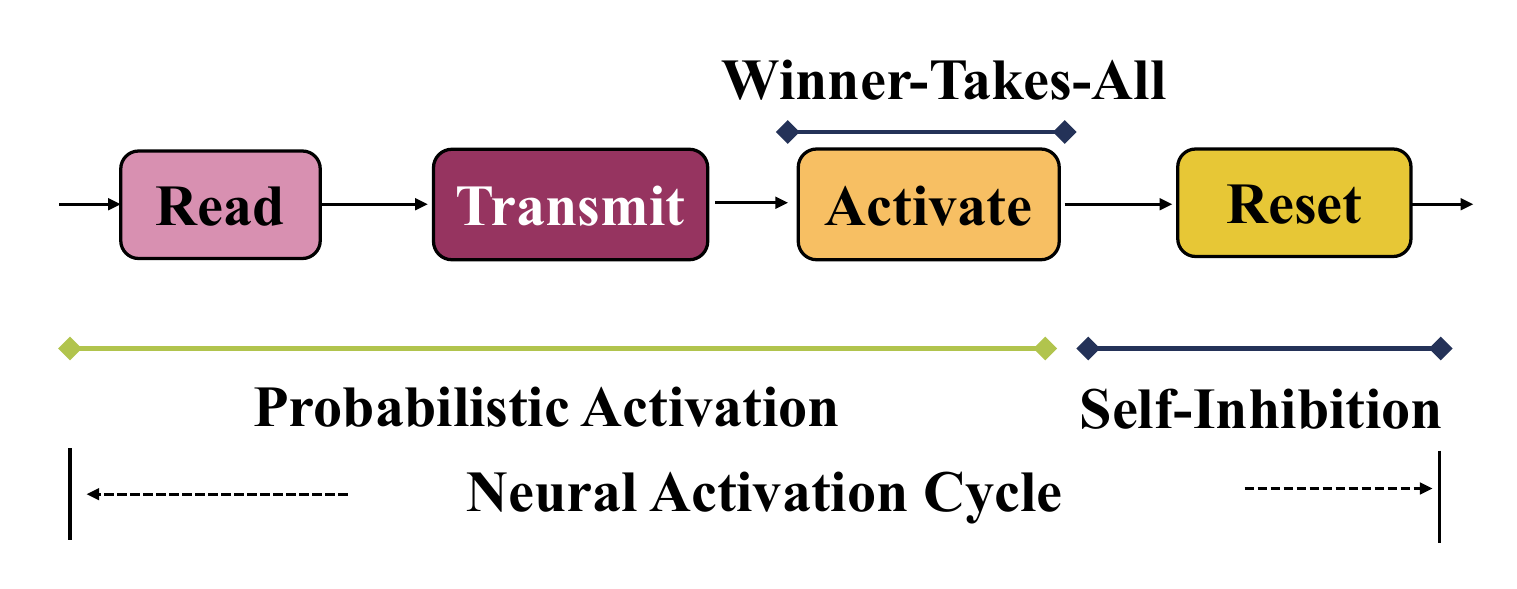}
\caption{A process of neural activation cycle.} 
\label{fig4.5}
\end{figure}

\begin{figure}[t]
\centering
\includegraphics[width=7cm]{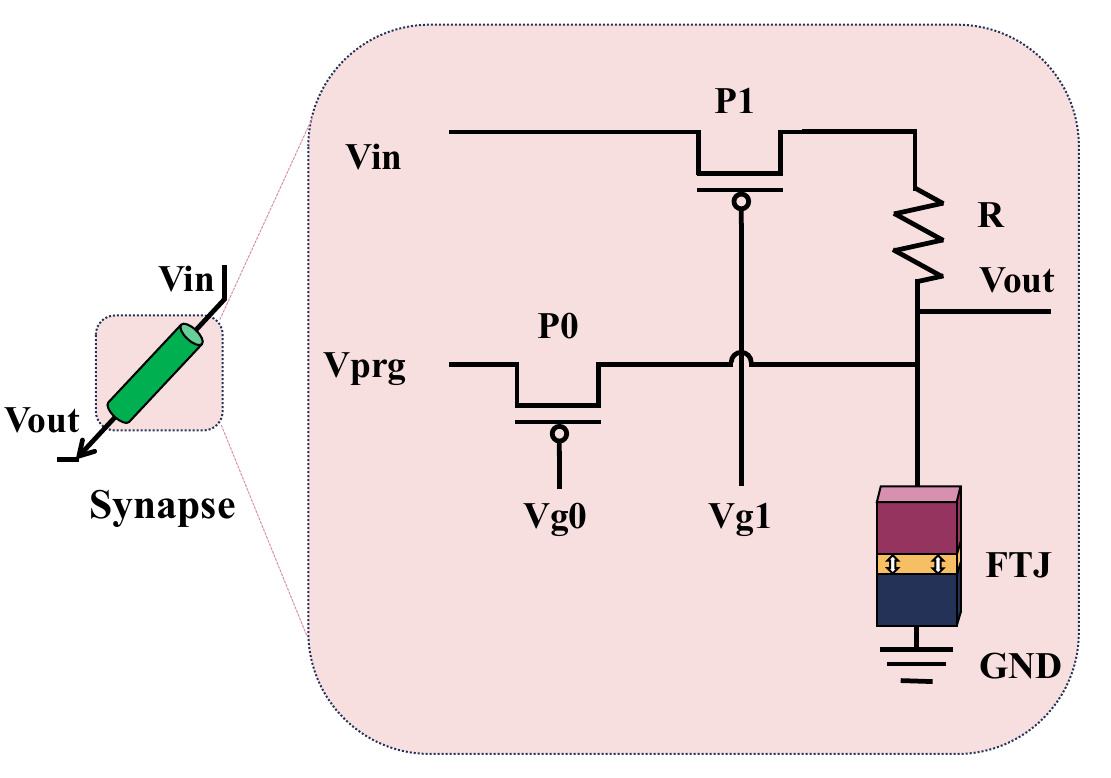}
\caption{NeuroPDE synaptic design: leveraging FTJ resistive switching to program synaptic weights.} 
\label{fig6}
\end{figure}
Fig.~\ref{fig4} illustrates the NeuroPDE architecture. Each neuron $\mathrm{N}_i$ represents a discrete position $\mathrm{X}_{i}$ in Fig.~\ref{fig1}, and the transition probability $P_{s}$ is stored in the synaptic weights $W$. Neurons are logically linear but physically arranged in odd/even rows for equal probability activation in both directions. Each neuron has five I/O pins: $\mathrm{in}$, $\mathrm{wr}$, $\mathrm{rd}$, $\mathrm{out}$, and $\mathrm{GND}$.



Fig.~\ref{fig4.5} illustrates the process of a single neural activation, with reference to the highlighted path in Fig.~\ref{fig4}. Each cycle is divided into two stages: probabilistic activation and self-inhibition. First, read the currently activated neuron, which is the $\mathrm{N}_{i+1}$ neuron in Fig.~\ref{fig4}, indicating that the walker is at position $\mathrm{X}_{i+1}$ in the one-dimensional discrete space. Then, multiply the read result by the weight $W$ at synapse $\mathrm{S}_{i+1}$. Next, attempt to activate neurons $\mathrm{N}_{i}$ and $\mathrm{N}_{i+2}$, allowing the walker to move left or right based on transition probability, with the winner-takes-all mechanism ensuring only one neuron is activated. After an activation attempt, if successful, the original neuron resets itself in a process known as self-inhibition. 
Repeating this process iteratively enables neural activation propagation through the circuit, simulating particle random walks.


\subsection{Synapse Design}

Fig.~\ref{fig6} illustrates an FTJ-based synapse, where $\mathrm{Vin}$ and $\mathrm{Vout}$ serve as input and output pins, respectively, linked to the $\mathrm{rd}$ of the presynaptic neuron and $\mathrm{wr}$ of the postsynaptic neuron, facilitating the reception of high voltage outputs from activated presynaptic neuron and attempting writes to two neighboring neurons. Synaptic has two modes of operation: programming and operational. Under programming mode, a low gate signal ($\mathrm{Vg0}$) opens transistor P0, allowing $\mathrm{Vprg}$ to be applied to the FTJ and programming its resistive state. In operational mode, a low gate signal ($\mathrm{Vg1}$) opens transistor P1, modulating $\mathrm{Vout}$ according to $\mathrm{R_{FTJ}}$, which means multiplying $\mathrm{Vin}$ by the weight $W$, as indicated in Equation (\ref{equSyn}). $\mathrm{Vout}$ affects the MTJ current, controlling the switching probability as shown in Equation (\ref{equMTJ}). The significantly lower $\mathrm{Vin}$ compared to $\mathrm{Vprg}$ ensures stable FTJ polarization. 
We will demonstrate the synaptic storage capability through SPICE simulations in Section~\ref{Exp:syn}.


\begin{equation}
\mathrm{Vout}=\frac{\mathrm{R_{FTJ}} }{\mathrm{R_{FTJ}}+\mathrm{R}} *\mathrm{Vin}.
\label{equSyn}
\end{equation}

\subsection{Neuron Design}

\begin{figure*}[t]
\centering
\includegraphics[width=17cm]{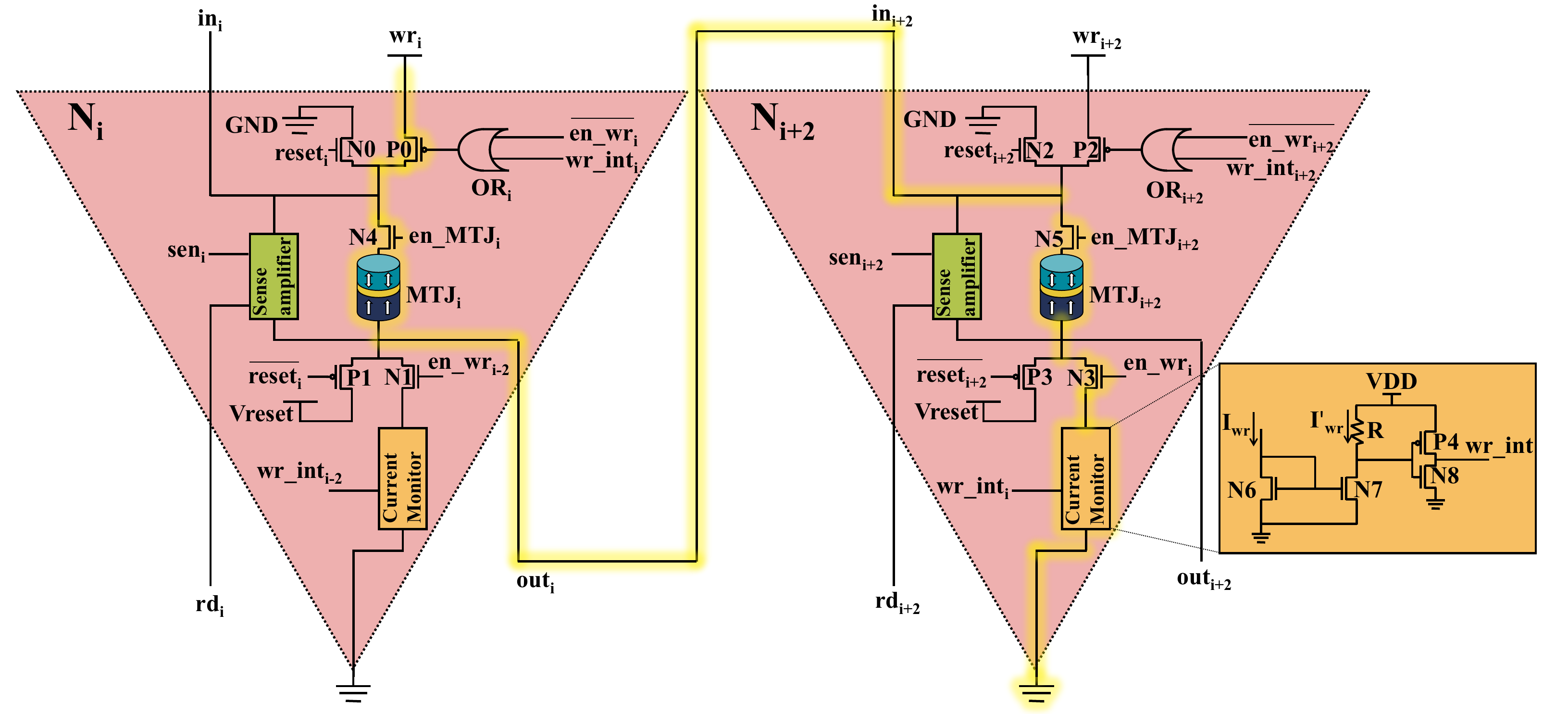}
\caption{NeuroPDE neuron design: integrating an MTJ for probabilistic activation, Sense Amplifiers for MTJ state detection, a Current Monitor to monitor MTJ switching, and supplementary read/write control circuits.} 
\label{fig7}
\end{figure*}

Fig.~\ref{fig7} shows the circuit designs of neurons, with pin arrangements consistent with those in Fig.~\ref{fig4}, focusing on illustrating their workings through examples of the left-neighboring neuron $\mathrm{N}_{i}$ and the right-neighboring neuron $\mathrm{N}_{i+2}$. We detail neuron functionality along the activation sequence. 

\textbf{Probabilistic Activation:} Initiating with the $\mathrm{sen}$ signal, the sense amplifier \cite{zhao2009high} is activated to detect the state of neurons, and subsequently conveys the MTJ status through $\mathrm{rd}$, where an active neuron outputs a high voltage to its synaptic destination. The output of the synapse will be fed into $\mathrm{wr}_{i}$. The OR value of the $\mathrm{\overline{en\_wr}}_{i}$ signal and $\mathrm{wr\_int}_{i}$ controls transistor P0, which is only enabled when both are low. Concurrently, transistors N3, N4, and N5 are enabled, facilitating the passage of the $\mathrm{wr}_{i}$ signal through $\mathrm{MTJ}_{i}$ to $\mathrm{out}_{i}$. The signal is then conveyed to the $\mathrm{in}_{i+2}$ of the neighboring neuron $\mathrm{N}_{i+2}$, following a path that includes $\mathrm{MTJ}_{i+2}$ and terminates at GND through the Current Monitor, as indicated by the highlighted path. This path essentially performs a random write operation on the two series-connected MTJs.

\textbf{Winner-Takes-All:} Upon an MTJ switch, resistance change triggers current variation in the $\mathrm{wr}_{i}$-$\mathrm{MTJ}_{i}$-$\mathrm{out}_{i}$-$\mathrm{in}_{i+2}$-$\mathrm{MTJ}_{i+2}$-GND path. The Current Monitor \cite{qu2018variation}, comprised of a current mirror and an inverter, detects and amplifies the current variation, raises $\mathrm{wr\_int}_{i}$, disabling P0 to interrupt writing. 

\textbf{Self-Inhibition:} Neural self-inhibition is accomplished through a controlled writing process. The $\mathrm{wr\_int}_{i}$ signal manages the $\mathrm{reset}$ signal, which in turn resets the MTJ of the initially activated neuron.

\section{Experiments and Evaluation}
\label{Experiments}

\subsection{Experimental Setup}

In the experiment, we conduct cycle-accurate transient simulations on circuits developed through SPICE simulations to evaluate their functionality. Additionally, we use extensive MC simulations with process variability to collect circuit-level neural activation histories, laying the foundation for system-level simulations. To facilitate the simulation of a large-scale neuron and synaptic circuit in a neuromorphic system solving PDEs through random walks, we employ a Python-based system-level simulator, which utilizes SPICE-derived activation histories to execute neuronal functions and facilitate successive random walks. We assess the performance of NeuroPDE by solving a thermal steady-state diffusion problem. Synaptic drift is inherently accounted for by modeling voltage variations in $V_{\mathrm{wr}_i}$ in Fig.~\ref{fig7} , obviating separate treatment in this study. This paper focuses on the single-thread performance of the architecture for tracking random walks, as the scalability of neuromorphic architectures for random walks has been studied \cite{smith2022neuromorphic}. 

Our circuit-level simulations are conducted utilizing the Cadence Virtuoso tools, incorporating the compact models of MTJ \cite{Wu2022} and FTJ \cite{wang2014compact}, along with the \textit{Cadence \SI{45}{\nano\meter} Generic Process Development Kit} (GPDK045). MC simulations incorporate MTJ stochasticity and $3\sigma$ process variations, with key MTJ parameters and variability strengths detailed in Table~\ref{tab1}, and primary FTJ parameters in Table~\ref{tab2}. All circuit-level simulations were conducted under an ambient temperature of \SI{300}{\kelvin}. Our system-level simulations can also use software-driven random walks to solve PDEs, enabling the evaluation of algorithm performance on general hardware as a benchmark for hardware evaluations. The NeuroPDE random walk simulator is developed in Python 3.7 with parallel computing support and runs on Ubuntu 20.04.1 with an Intel\textsuperscript{\textregistered} Core\texttrademark\ i9-12900 CPU.

\begin{table}[t]
\centering
\caption{Key device parameters for MTJ compact model.}\label{tab1}
\resizebox{0.45\textwidth}{!}{
\begin{tabular}{
>{}c |
>{}c |
>{}c }
\hline
{ \textbf{Parameter}} & {\textbf{Description}}  & {\textbf{Value}}  \\ \hline \hline
{$t_{\mathrm{FL}}$}       & {Thickness of the free layer}        & {1.3nm}  \\ \hline
{$\sigma _{t_{\mathrm{FL}}} $}       & {Standard deviation of $t_{\mathrm{FL}}$}        & {3\% of 1.3nm}  \\ \hline

{$CD$}         & {Critical diameter}        & {32nm}   \\ \hline
{$t_{\mathrm{TB}}$}       & {Thickness of the tunnel barrier}     & {0.85nm} \\ \hline

{$\sigma _{t_{\mathrm{TB}}} $}       & {Standard deviation of $t_{\mathrm{TB}}$}     & {3\% of 0.85nm} \\ \hline

{$TMR$}       & {TMR ratio} & {200\%}   \\ \hline
{$\sigma _{TMR}$}       & {Standard deviation of TMR} & {3\% of 200\%}   \\ \hline
\end{tabular}
}
\end{table}

\begin{table}[t]
\centering
\caption{Key device parameters for FTJ compact model.}\label{tab2}
\resizebox{0.5\textwidth}{!}{
\begin{tabular}{
>{}c |
>{}c |
>{}c }
\hline
{ \textbf{Parameter}} & {\textbf{Description}}  & {\textbf{Value}}  \\ \hline \hline
{$t_{B}$}       & {Barrier thickness}        & {2nm}  \\ \hline
{$r$}        & {Junction surface radius}  & {175nm}   \\ \hline
{$U_{\mathrm{N}}$} & {Creep energy barrier for the domain} & {0.67eV} \\ \hline
{$U_{\mathrm{P}}$} & {Creep energy barrier for the domain wall} & {0.52eV} \\ \hline
{$\tau_{\mathrm{0N}}$} & {Attempt time of the domain nucleation} & {2.8e-15s} \\ \hline
{$\tau_{\mathrm{0P}}$} & {Attempt time of the domain wall} & {9e-14s} \\ \hline
{$\varphi_{\mathrm{1OFF}}$} & {Barrier potential height at
LSMO/BTO interface(OFF)} & {0.678V} \\ \hline
{$\varphi_{\mathrm{1ON}}$} & {Barrier potential height at
LSMO/BTO interface(ON)} & {0.53V} \\ \hline
{$\varphi_{\mathrm{2OFF}}$} & {Barrier potential height at
Co/BTO interface(OFF)} & {0.978V} \\ \hline
{$\varphi_{\mathrm{2ON}}$} & {Barrier potential height at
Co/BTO interface(ON)} & {1.014V} \\ \hline
{$m_{\mathrm{OFF}}$} & {Effective electron mass(OFF)} & {0.931me} \\ \hline
{$m_{\mathrm{ON}}$} & {Effective electron mass(ON)} & {0.437me} \\ \hline

\end{tabular}
}
\end{table}

\subsection{Circuit-Level Simulation}

\subsubsection{Synaptic Function Evaluation}
\label{Exp:syn}

\begin{figure}[t]
\centering
\includegraphics[width=9cm]{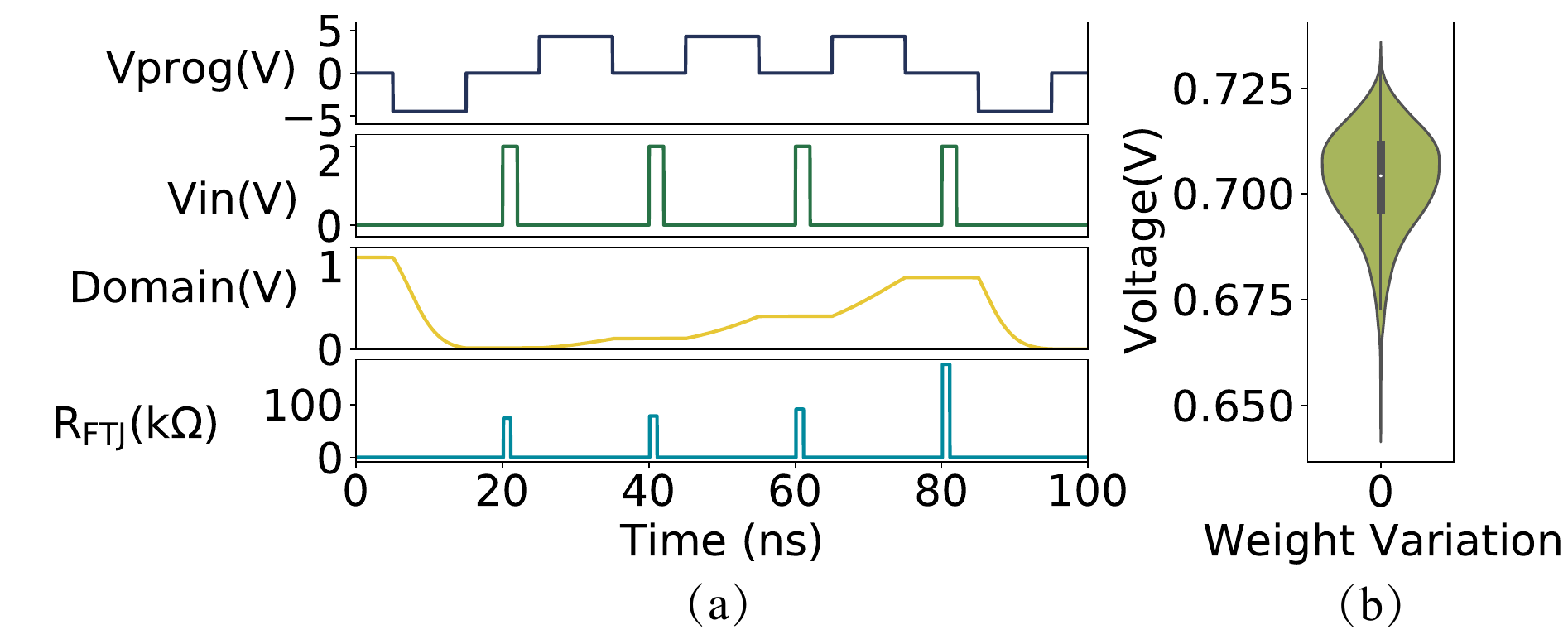}
\caption{(a) Transient simulation results for synapses. (b) MC simulation results for synapses.}
\label{fig9}
\end{figure}

Fig.~\ref{fig9}(a) presents the SPICE simulation results for a synapse based on the FTJ. The figure sequentially depicts from top to bottom: the programming voltage ($\mathrm{Vprg}$), the activation voltage corresponding to the synaptic input voltage ($\mathrm{Vin}$), the polarization direction of the ferroelectric domain walls (Domain), and the FTJ resistance ($\mathrm{R_{FTJ}}$). 

In the programming mode
, a negative $\mathrm{Vprg}$ pulse between \SI{5}{\nano\second}--\SI{15}{\nano\second} switches the FTJ to the fully ON state. Subsequently, positive pulses are applied in intervals of \SI{25}{\nano\second}--\SI{35}{\nano\second}, \SI{45}{\nano\second}--\SI{55}{\nano\second}, and \SI{65}{\nano\second}--\SI{75}{\nano\second} to program the FTJ, facilitating consecutive domain growth. Between \SI{85}{\nano\second}--\SI{95}{\nano\second}, another negative $\mathrm{Vprg}$ pulse resets the domain walls. This sequence demonstrates the bidirectional growth of domains under programmed pulse control, thereby illustrating the memristive characteristics of the FTJ. In the operational mode, 
read pulses are injected at intervals of \SI{20}{\nano\second}, \SI{40}{\nano\second}, \SI{60}{\nano\second}, and \SI{80}{\nano\second} with a small amplitude $\mathrm{Vin}$. These pulses do not alter the domain polarization, indicating stability of the synapse design. Varying domain polarizations result in different resistances, allowing each read pulse to detect unique FTJ states.

Fig.~\ref{fig9}(b) shows the voltage distribution of synaptic weights after \SI{15}{\nano\second} of programming in 50,000 Monte Carlo simulations. The weight distribution is non-normal, with a mean shift of less than 0.32\% and variance below 0.000138. We will consider the impact of weight precision variations on the system simulation.

\subsubsection{Neuron Activation Evaluation}

\begin{figure}[t]
\centering
\includegraphics[width=9cm]{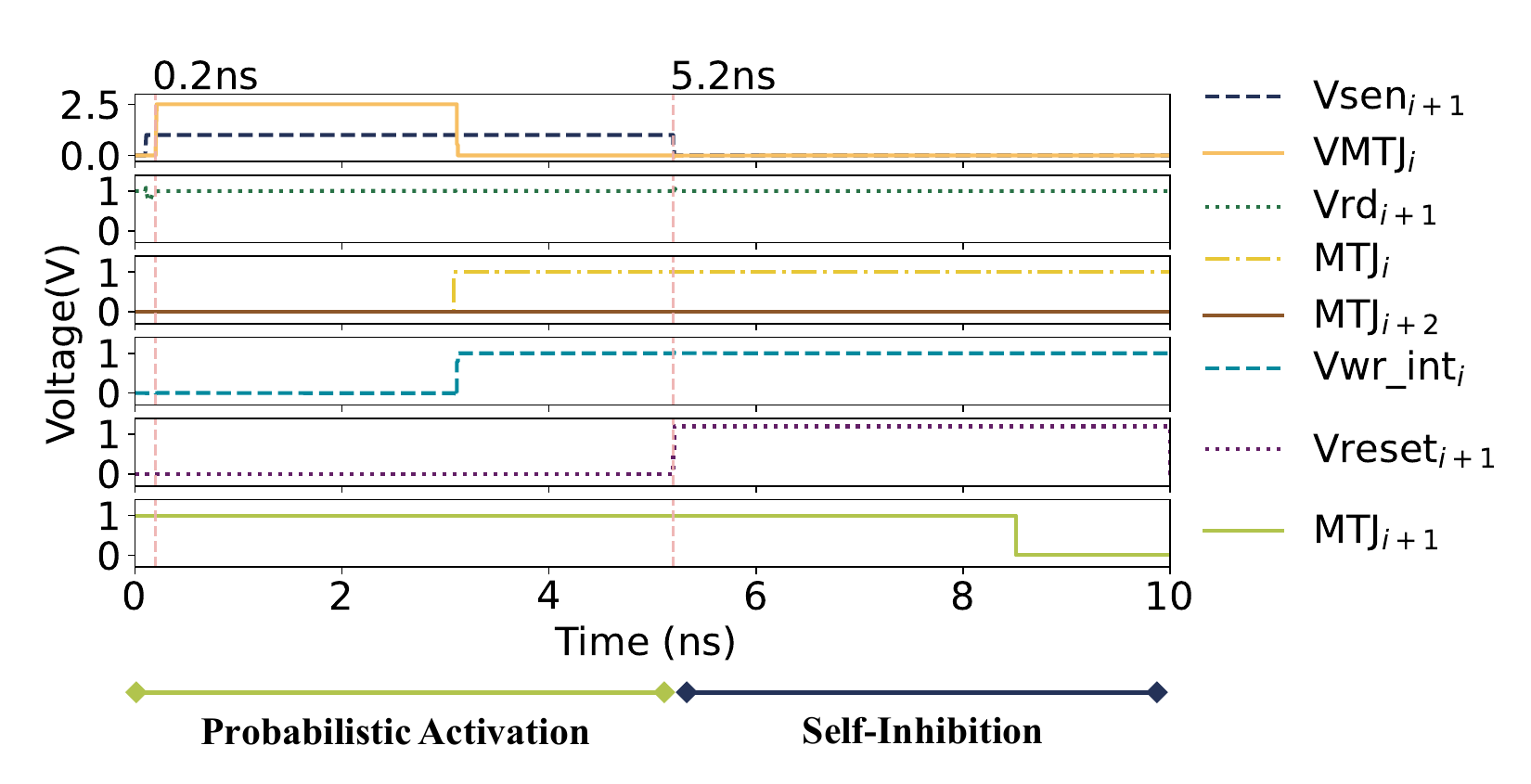}
\caption{The transient simulates probabilistic activation of neurons $\mathrm{N}_{i}$ and $\mathrm{N}_{i+2}$ by $\mathrm{N}_{i+1}$, with the figure illustrating the activation process of neuron $\mathrm{N}_{i}$.} 
\label{fig10}
\end{figure}

Fig.~\ref{fig10} illustrates a neuronal activation event within a cluster involving presynaptic neuron $\mathrm{N}_{i+1}$, and its adjacent postsynaptic neurons $\mathrm{N}_{i}$ and $\mathrm{N}_{i+2}$. Activation pathways are highlighted in Fig.~\ref{fig4}, while neuronal circuit details are provided in Fig.~\ref{fig7}. Activation commences with the $\mathrm{Vsen}_{i+1}$ signal triggering the sense amplifier in neuron $\mathrm{N}_{i+1}$, which precharges and read in \SI{0.2}{\nano\second} and transmits a signal of ``1" via $\mathrm{Vrd}_{i+1}$. This weighted signal, via synapses, becomes $\mathrm{Vwr}_{i}$  for neuron $\mathrm{N}_{i}$. Within the next \SI{5}{\nano\second}, $\mathrm{Vwr}_{i}$ is applied through P0 to the source of N4, applying $\mathrm{VMTJ}_{i}$ to attempt to write on $\mathrm{MTJ}_{i}$ and $\mathrm{MTJ}_{i+2}$. In this simulation, $\mathrm{MTJ}_{i}$ switches, resulting in increased resistance and reduced current. The Current Monitor of neuron $\mathrm{N}_{i+2}$ responds by raising its output ($\mathrm{Vwr\_int}_{i}$) and pulling down $\mathrm{VMTJ}_{i}$, ensuring $\mathrm{MTJ}_{i+2}$ remains unchanged, implementing a winner-takes-all mechanism. Following this, self-inhibition occurs, suppressing the activation of the original neuron. The $\mathrm{Vwr\_int}_{i}$ signal indicates the activation event and controls $\mathrm{Vreset}_{i+1}$ to output a ``1" signal during the self-inhibition phase, ensuring that $\mathrm{MTJ}_{i+1}$ is reset within the required time. Additional MC simulations, integrating process variations and MTJ switching randomness, produce a table of neuron activation histories from 50,000 simulations. This table is subsequently used in system-level simulations.


\vspace{-5pt} 

\subsection{System-Level Experiment}

\subsubsection{1D Steady-State Heat Equation Problem}

In this section, we explore a steady-state problem with boundary conditions. We present a specific example of a \textit{one-dimensional} (1D) steady-state heat equation: A thin metal wire of length $L$  has one end at $x=0$ exposed to an external temperature $T=0$, and the thermal gradient is zero. At the opposite end, there is a heat source with a gradient of $-F$, diminishing linearly towards the left endpoint. The steady-state temperature distribution along the wire at position $x$, denoted by $u(x)$, is given by:
\vspace{-5pt} 
\begin{equation}
    \begin{split}
   &0 =\frac{\mathrm{d^{2} } }{\mathrm{d} x^{2} } u-F(L-x),x\in [0,L],\\
          &u(0) = 0,  \quad
       {u}'(0) = 0.
    \end{split}
\label{equ2}
\end{equation}

This problem has an analytical solution as follows:
\vspace{-5pt} 
\begin{equation}
u(x)=\frac{FLx^2}{2}- \frac{Fx^3}{6}.
\end{equation}

\begin{figure*}[t]
\centering
\includegraphics[width=13cm]{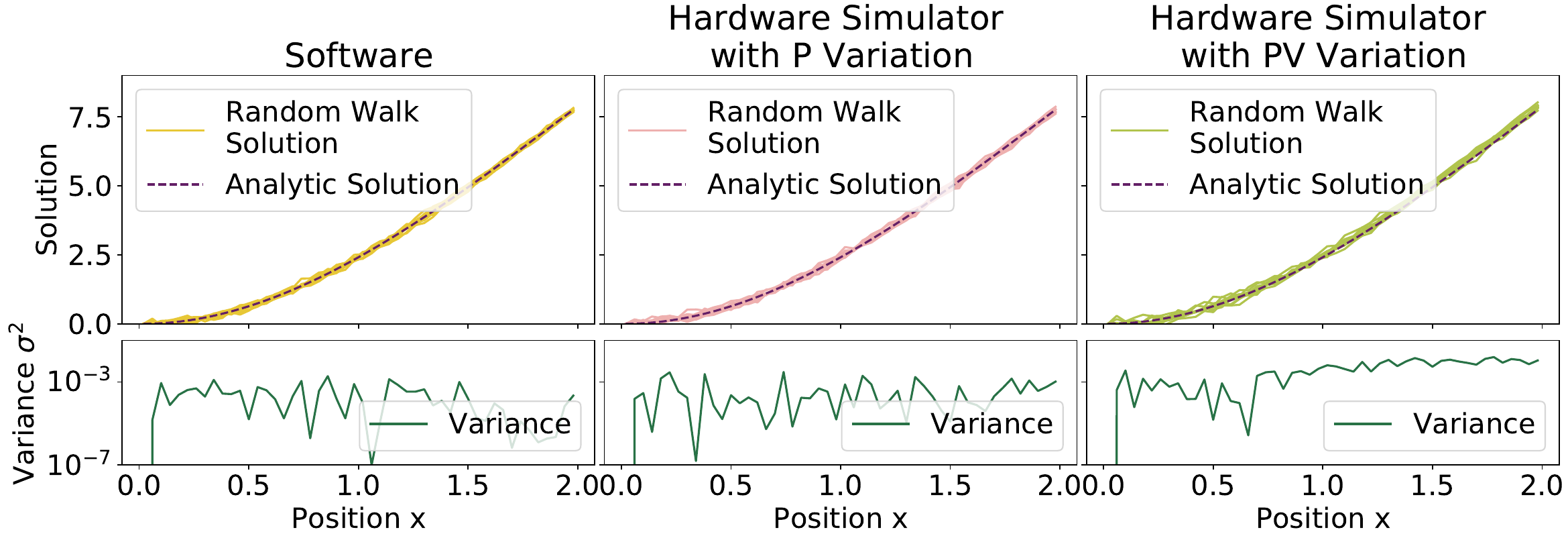}
\caption{Comparison of variances in solving the steady-state problem between software-based random walk and its hardware counterpart under the impact of process and voltage variations.
} 
\label{fig13}
\end{figure*}

\begin{figure*}[t]
\centering

\includegraphics[width=0.8\textwidth]{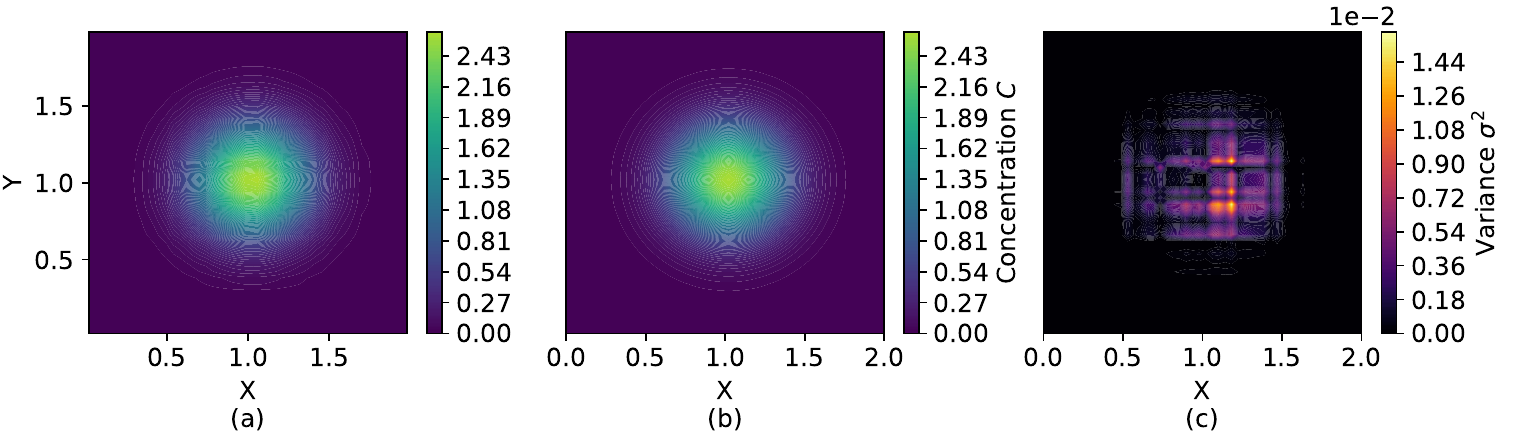}
\caption{Comparison of variance between NeuroPDE solution and analytical solution for the 2D diffusion equation.} 
\label{figaddmc} 
\end{figure*}

\subsubsection{Modeling and Solving}

Next, we describe how to solve this equation using the MC random walk method. We construct a Markov chain to execute random walks. Detailed information can be found in this work \cite{smith2020solving}.

\begin{itemize}
    \item A 1D space of length $L$ is discretized into $N$ positions, with $W$ walkers initialized at each position.
    \item Simulate particles among these $W$walkers: each step takes a time $dt$. The particle has a probability $P_{s}$ of staying at the same position, and a probability $P_{g}$ for moving left or right. At the position $x=0$, the probability of moving right is $2P_{g}$.
    \item Track each walker until it reaches $x=L$ and then terminate the simulation. Record the initial position $\mathrm{X}_i$ and count the number of passages through $\mathrm{X}_j$ as $n_{i,j}$, forming a matrix documenting walking histories.
    \item Calculate the solution to the PDE using Equation (\ref{equux}).
\end{itemize}

\begin{equation}
    \begin{split} 
    &\mathbb{E} [-F\int_{0}^{T}L-X(s)\,\mathrm{d}s  \mid X_{0} = \mathrm{X}_{i}] \\
    &\approx -\frac{F\cdot dt}{W}\sum_{j}n_{i,j}(l-\mathrm{X}_{j}):= u_{i}.\\
    &u(\mathrm{X}_{i}) \approx u_{i}-u_{0}.
    \end{split}
\label{equux}
\end{equation}

The values of probability ($P_{g}$) and ($P_{s}$) correspond to the integrals of the probability density function (PDF) over discrete states, as given by Equation (\ref{equP}) in the diffusion problem.

\begin{equation}
\begin{aligned}
P_s & = \mathbb{P}\left[-\frac{1}{2}\Delta x < X_{\Delta t} < \frac{1}{2} \Delta x \mid X_0 = \mathrm{X}_i \right] \\
& = \frac{1}{2\sqrt{\Delta t \pi}} \int_{-\frac{1}{2}\Delta x}^{\frac{1}{2}\Delta x} \exp \left( -\frac{x^2}{4\Delta x} \right) \mathrm{d}x, \\
P_g & = \mathbb{P}\left[X_{\Delta t} \leq -\frac{1}{2} \Delta x \mid X_0 = \mathrm{X}_i \right] \\
& = \frac{1}{2\sqrt{\Delta t \pi}} \int_{-\infty}^{-\frac{1}{2} \Delta x} \exp \left( -\frac{x^2}{4\Delta x} \right) \mathrm{d}x.
\end{aligned}
\label{equP}
\end{equation}

The parameter values in our simulations are $L=2$, $N=50$ $dt=0.00038$, $F=3$ and $W=1e4$. Fig.~\ref{fig13} shows three PDE solutions: software-driven random walks, hardware simulations with process (P) variations, and with process and voltage (PV) variations. The voltage variation originates from the synaptic weight distribution shown in Fig.~\ref{fig9}(b), and is applied to
$V_{\mathrm{wr}_i}$ in Fig.~\ref{fig7} to introduce voltage variation in the system simulation. This results from synaptic drift and precision errors induced by weight updates. The dashed line represents the analytical solution ($C_{\mathrm{an}}$), while the 10 iterations of random walk solutions ($C_{\mathrm{rw}}$) using MC random walks are shown nearby, with variance $\sigma ^{2}$ calculated with the following equation: 
\begin{equation}
\sigma ^{2}_i = \left | \overline{C_{\mathrm{rw}} (\mathrm{X}_{i})}   -C_{\mathrm{an}}(\mathrm{X}_{i})\right | ^{2}.
\label{variation}
\end{equation}


It can be seen that our NeuroPDE design effectively resolves PDEs, with simulated hardware results showing a variance  below 1e-3 when considering process variation, which is consistent with the software-based solution. 
Introducing voltage offsets from synaptic precision variations slightly increases the variance as $x$ nears $L$, due to a slight reduction in the mean of the synaptic weight distribution from precision and drift. Meanwhile, the nonlinear dependence of walk probability on activation voltage lowers the expected activation probability, increasing the solution value. However, the large number of walks ensures error tolerance, with variance remains below 1e-2 highlighting the tolerance of NeuroPDE to voltage variations.

This simplified diffusion problem has broad implications. By leveraging the strength of the MC method in high-dimensional problems, NeuroPDE tracks a single variable across a single unit, enabling the solution of more complex problems. Fig.~\ref{figaddmc} presents the solution to the \textit{two-dimensional} (2D) diffusion equation Equation (\ref{equ2V}), where two sets of NeuroPDE units are employed to perform random walk tracking for the X and Y variables. The hardware tracking and solving process is similar to the 1D steady-state problem, with the key difference being that each walker is initialized at the point source. At time $t$, the position vectors $\mathbf{u}_{i}$ of all tracked particles are recorded to compute the single-dimension solution $\mathbf{u}(\mathrm{X}_{i})$, while independent tracking of two variables yields the two-dimensional solution $C(X,Y,t)= \mathbf{u}(\mathrm{X}_i) \cdot \mathbf{u}(\mathrm{Y}_j)$. 


With parameters set to $c0=1$, $D=1$, $W=1e5$, $t=80dt$, and other conditions consistent with the 1D steady-state problem, the hardware solution (Fig.~\ref{figaddmc}(a)) achieves a variance of less than 1e-2 compared to the analytical solution (Fig.~\ref{figaddmc}(b)).
  
\begin{equation}
\left\{
\begin{array}{ll}
\frac{\partial C(X, Y, t)}{\partial t} = D \left( \frac{\partial^2 C(X, Y, t)}{\partial X^2} + \frac{\partial^2 C(X, Y, t)}{\partial Y^2} \right), \\
C(1, 1, 0) = c0.
\end{array}
\right.
\label{equ2V}
\end{equation}

\begin{figure}[t]
\centering
\includegraphics[width=0.48\textwidth]{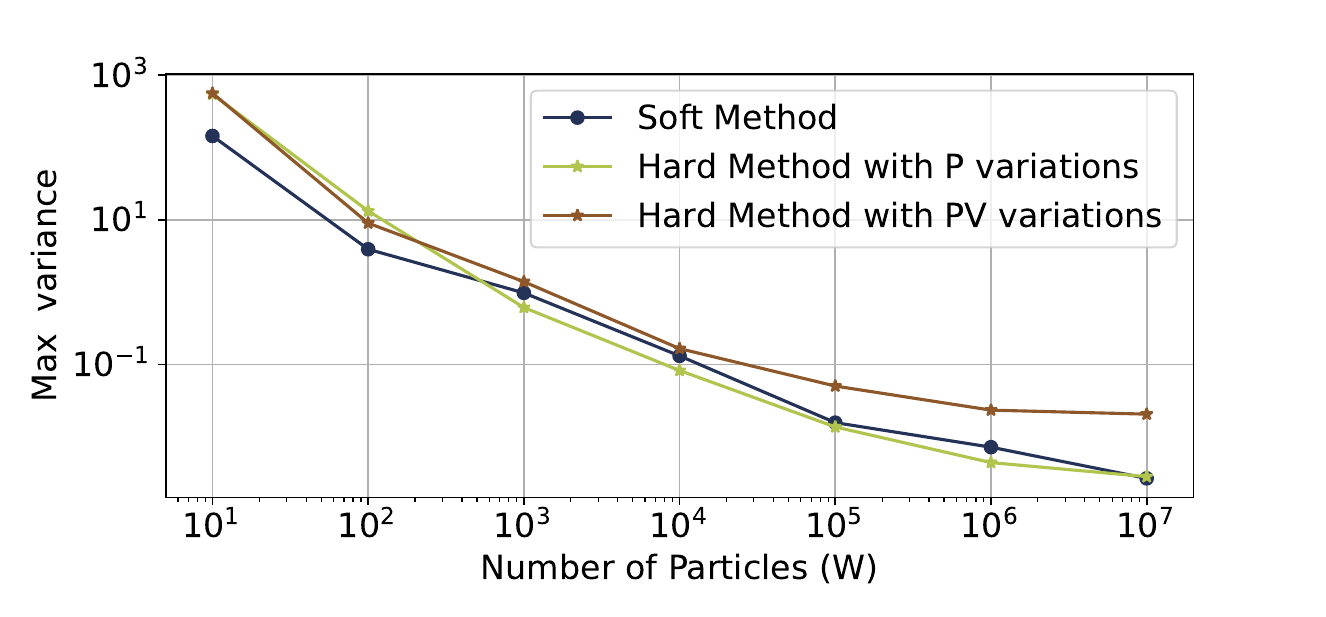}
\caption{Variation in maximum variance between the random walk and analytical solutions, comparing software methods and NeuroPDE hardware solutions, as a function of particle scale.} 
\label{figaddW} 
\end{figure}

Fig.~\ref{figaddW} further investigates the variance as a function of particle scale in solving the 2D diffusion problem. The horizontal axis represents the particle scale in MC simulations, while the vertical axis denotes the maximum variance of the random walk solution relative to the analytical solution. Comparisons are made between the software method and the NeuroPDE hardware methods, considering P variations and PV variations. Results show that without voltage variations, the NeuroPDE method exhibits no significant difference from the software random walk. However, introducing voltage variation increases the particle scale and reduces the maximum variance by approximately one order of magnitude at convergence, which is consistent with our evaluation of the 1D steady-state problem in Fig.~\ref{fig13}.

\subsubsection{Performance and Energy}

\begin{figure}[t]
\centering
\includegraphics[width=0.49\textwidth]{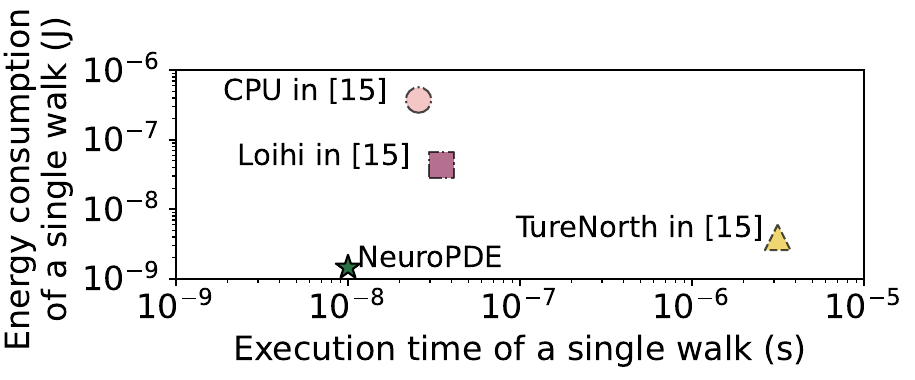}
\caption{Comparison of NeuroPDE with a single-core CPU, Loihi and TrueNorth \cite{smith2022neuromorphic} in terms of performance and energy consumption for executing a single MC random walk.
} 
\label{fig14}
\end{figure}

Fig.~\ref{fig14} shows the performance and energy improvements of NeuroPDE in comparison to random walks on other architectures. We evaluate performance and energy consumption based on the execution time and consumed energy of a single walk. NeuroPDE completes each walk in \SI{10}{\nano\second} with a energy consumption of \SI{1.451}{\pico\joule}, according to SPICE simulations. We compared our design with the state-of-the-art neuromorphic PDE solvers \cite{smith2022neuromorphic}. Compared to prior neuromorphic chips, NeuroPDE achieves a 3.48$\times$ to 315$\times$ speedup in execution time while offering 2.7$\times$ to 29.8$\times$ higher energy efficiency (reducing energy to 3.35\% to 36.7\% of prior neuromorphic PDE solvers).

In summary, CMOS-based neuromorphic architectures such as \cite{smith2022neuromorphic} excel in energy consumption over the von Neumann architecture but lack performance benefits due to limited inherent randomness. By utilizing the physical randomness of emerging spintronic devices, we can further boost the performance  and energy efficiency of neuromorphic architectures as demonstrated with our NeuroPDE design.

\section{Discussion}
\label{Discussion}

Table~\ref{tab5} compares different types of PDE solvers. \textbf{The Neural Network Solver} is limited by its training overhead, preventing it from matching the performance of the unsupervised NeuroPDE. \textbf{Prior MC methods} offer dimensional flexibility and leverage parallel processing capabilities. However, as stochastic methods, they inherently introduce discrepancies to the conventional deterministic architectures, affecting computational efficiency. \textbf{The proposed NeuroPDE design} incorporates physical randomness at the device level, effectively bridging the gap between  algorithms and the underlying  architectural design  for solving PDEs.

\begin{table}[t]
\centering
\caption{Comparison between various PDE solvers.}\label{tab5}
\resizebox{0.488\textwidth}{!}{
\begin{tabular}{p{4.2cm}lllll}

\hline

\large\textbf{PDE Solver Type} & \large\begin{tabular}[c]{@{}l@{}}\textbf{Problems} \\\textbf{without} \\ \textbf{Analytical}\\ \textbf{Solutions}\end{tabular} &\large \begin{tabular}[c]{@{}l@{}}\textbf{Problems} \\\textbf{with} \\ \textbf{High-}\\ \textbf{dimensionality}\end{tabular} &\large \begin{tabular}[c]{@{}l@{}}\textbf{No} \\\textbf{Training} \\  \textbf{Overhead}\end{tabular} &\large \begin{tabular}[c]{@{}l@{}}\textbf{High} \\ \textbf{Performance}\end{tabular}\\

\hline \hline
\large Analytical Solver \cite{henner2019partial}&\XSolid{}&\XSolid{}&\Checkmark{}&NA\\ 
\large Numerical Solver \cite{Claes1990,townsend2015automatic,langtangen2017finite}&\Checkmark{}&\XSolid{}&\Checkmark{}&\XSolid{}\\ 
\large Neural Network Solver \cite{blechschmidt2021three,karniadakis2021physics,meng2023pinn, Pestourie2023} &\Checkmark{}&\Checkmark{}&\XSolid{}&\XSolid{}\\ 
\large Prior MC Random Walk Solver  \cite{severa2018spiking, smith2020solving, smith2022neuromorphic,zhang2023monte}&\Checkmark{}&\Checkmark{}&\Checkmark{}&\XSolid{}\\ 
\large NeuroPDE (this work)&\Checkmark{}&\Checkmark{}&\Checkmark{}&\Checkmark{}\\\hline
\end{tabular}}
\vspace{-5pt} 
\end{table}

The current design prioritizes solving PDEs under PV variations, as temperature-induced probabilistic drift primarily impacts performance in complex operating environments and long-term stability. This represents a universal challenge in existing non-volatile memory-based neuromorphic computing systems \cite{torres2023thermal}, and also lies beyond the scope of this study, which aims to evaluate feasibility of PDE solvers with hardware stochasticity. Additionally, broader applicability and solutions for high-dimensional complex problems need extended consideration. Notably, the theoretical foundation for high-dimensional scalability remains explicit, where the intrinsic dimension-agnostic convergence of MC methods theoretically supports solving complex systems. Future research should primarily focus on: 1) multidimensional topological neuron arrays required for high-dimensional extension, 2) error-correction mechanism design with emphasis on temperature compensation, and 3) collaboration with domain experts on practical problems to consolidate practical value of NeuroPDE. 



\section{Conclusion}
\label{Conclusion}

This paper introduces a neuromorphic PDE solver using spintronic and ferroelectric devices. It simulates random walks through neural activations on a neuromorphic system to solve PDEs effectively. Our design features neurons with probabilistic activation, winner-takes-all, and self-inhibitory functions, and synapses that can store weights continuously. We evaluated the functionality of NeuroPDE through SPICE simulations and implemented a system simulation to solve the steady-state heat equation. While ensuring solution accuracy, we achieved significant performance and energy efficiency advantages compared to both general-purpose CPUs and CMOS-based neuromorphic chips. By leveraging device-level randomness, we close the gap between computing architecture with probabilistic algorithms, overcoming the computational overhead of conventional deterministic systems. This  paves the way for using neuromorphic circuits in scientific computations.

\balance
\bibliographystyle{IEEEtran}

\bibliography{ref}



\balance

\vfill
\end{document}